\documentclass[structabstract]{aa} 
\usepackage{natbib}
\usepackage{graphicx}
\usepackage{longtable}
\usepackage{amsmath} 
\usepackage{units}

\newcommand{\kms}{$\mathrm{km\, s^{-1}\, }$}
\newcommand{\mbh}{M_{\bullet}}

\newcommand{\msun}{{\rm M}_{\odot}}

\usepackage{txfonts}
\begin{document}
   \title{The $\mbh - \sigma$ relation for intermediate-mass black holes in globular clusters}
   
   \titlerunning{The $\mbh - \sigma$ relation for intermediate-mass black holes}

   \author{N.~L\"utzgendorf
          \inst{1}
          \and
          M.~Kissler-Patig\inst{2}
          \and
          N.~Neumayer\inst{1}
          \and
          H.~Baumgardt\inst{3}          
          \and
          E.~Noyola\inst{4,5}
		  \and
          P.~T.~de Zeeuw\inst{1,6}
          \and
          K.~Gebhardt\inst{7} 
          \and
          B.~Jalali\inst{8}
          \and
          A.~Feldmeier \inst{1}
          }

   \institute{European Southern Observatory (ESO),
              Karl-Schwarzschild-Strasse 2, 85748 Garching, Germany\\
              \email{nluetzge@eso.org}
         \and     
              Gemini Observatory, Northern Operations Center, 670 N. A'ohoku Place
Hilo, Hawaii, 96720, USA
         \and
			 School of Mathematics and Physics, University of Queensland, 
			 Brisbane, QLD 4072, Australia
         \and
             Instituto de Astronomia, Universidad Nacional Autonoma de Mexico (UNAM), 
             A.P. 70-264, 04510 Mexico
             %\email{c.ptolemy@hipparch.uheaven.space}
         \and
			 University Observatory, Ludwig Maximilians University, 
			 81679 Munich, Germany	 
         \and
			 Sterrewacht Leiden, Leiden University, 
			 Postbus 9513, 2300 RA Leiden, The Netherlands
         \and
			 Astronomy Department, University of Texas at Austin, 
			 Austin, TX 78712, USA 
         \and
	         I.Physikalisches Institut, Universit\"at zu K\"oln, 
    	     Z\"ulpicher Str. 77, 50937 K\"oln, Germany}

   \date{Received January 29, 2013; accepted May 26, 2013}

% \abstract{}{}{}{}{} 
% 5 {} token are mandatory
 
  \abstract
   {For galaxies hosting supermassive black holes (SMBHs), it has been observed that the mass of the central black hole ($\mbh$) tightly correlates with the effective or central velocity dispersion ($\sigma$) of the host galaxy. The origin of this $\mbh - \sigma$ scaling relation is assumed to lie in the merging history of the galaxies but many open questions about its origin and the behavior in different mass ranges still need to be addressed.}
   {The goal of this work is to study the black-hole scaling relations for low black-hole masses, where the regime of intermediate-mass black holes (IMBHs) in globular clusters (GCs) is entered.}
   {We collect all existing reports of dynamical black-hole measurements in globular clusters, providing black-hole masses or upper limits for 14 candidates. We plot the black-hole masses versus different cluster parameters including total mass, velocity dispersion, concentration and half-mass radius. We search for trends and test the correlations in order to quantify their significance using a set of different statistical approaches. For correlations showing a large significance we perform a linear fit, accounting for uncertainties and upper limits.}
   {We find a clear correlation between the mass of the IMBH and the velocity dispersion of the globular cluster. As expected, the total mass of the globular cluster then also correlates with the mass of the IMBH. While the slope of the  $\mbh - \sigma$ correlation differs strongly from the one observed for SMBHs, the other scaling relations $M_{\bullet} - M_{tot}$, and $M_{\bullet} - L$ are similar to the correlations in galaxies. Significant correlations of black-hole mass with other cluster properties were not found in the present sample.}
   % ty if necessary 
{}


   \maketitle
%
%__________________________________________________________________

\section{Introduction}

\subsection{The low-mass end of the $\mbh - \sigma$ relation}

Empirical scaling relations between the masses of nuclear black holes and properties of their host galaxies in terms of total luminosity L, bulge mass $M_{\scriptsize{\mbox{bulge}}}$ \citep{Kormendy_1995, Marconi_2003, haering_2004} or the effective (projected) velocity dispersion $\sigma$ \citep[][]{ferrarese_2000, Gebhardt_2000, ferrarese_2005, gultekin_2009} disclose a strong connection between the formation history of supermassive black holes (SMBHs) and galaxies. These scaling relations can be explained  by gas accreting black holes which drive powerful jets and outflows into the surrounding interstellar medium, thus affecting the star formation rate and regulating the further supply of matter onto the SMBH \citep{silk_1998, Matteo_2005, Springel_2005, Hopkins_2007}. Alternative explanations reproduce these empirical correlations by large numbers of merger events in which extreme values of $\mbh/M_{\scriptsize{\mbox{bulge}}}$ are averaged out. 

While galaxies and central SMBHs in the most common mass range, $\mbh=\unit{10^{6-9}}{\msun}$, seem to closely follow the one parameter power law $\mbh-\sigma$ and $\mbh-L$ relations, there is now some evidence for large scatter or upward curvature trend for SMBH masses at the highest mass regime \citep[$\mbh\approx\unit{10^{10}}{\msun}$, e.g.][]{Gebhardt_2011,McConnell_2011,bosch_2012}. 

\begin{table*}\tiny
\caption{Properties of the 14 globular clusters of our sample. The references for the measurements are: a) \cite{harris_1996}, b) \cite{nora11}, c) \cite{nora12a}, d) \cite{nora13} e) \cite{feldmeier_2013}, f) \cite{gerssen_2002}, g) \cite{McLaughlin_2005}, h) \cite{ma_2007} i) \cite{meylan_2001} j) \cite{stephens_2001}, k) \cite{noyola_2010}, l) \cite{ibata_2009}, m) \cite{gebhardt_2005} n) \cite{meylan_1995}. For those clusters not provided with uncertainties in $M_{tot}$ and $L_{tot}$ we adopted a value of $0.02$ dex. }             % title of Table
\label{tab:prop}      % is used to refer this table in the text
\centering
\begin{tabular}{lcccccccccccccc}
\hline \hline
\noalign{\smallskip}
ID && $D_{SUN}$ & $D_{GC}$ & [Fe/H] & e & c & $r_h$ & $\log M_{tot}$& $\delta M_{tot}$& $\log L_{tot}$& $\delta L_{tot}$ & $\sigma$ & $\delta \sigma$ & $M_{\bullet}$\\
   & &[kpc]     & [kpc]    &        &   &    & [pc]& $[\log M_{\odot}]$& & $[\log L_{\odot}]$ &  & [km s$^{-1}]$ & & $[M_{\odot}]$\\
 \noalign{\smallskip}
\hline
\noalign{\smallskip}
       G1  & 				&$675.0^i$ & $ 36.8^j$ & $-1.22^h$ & $ 0.19^h$ & $ 2.01^h$ & $ 6.5^h$ & $6.76^m$ & $0.02$ & $6.31^a$ & $0.02$ & $25.1^i$ & $ 0.3$ & $	( 1.8 \pm  0.5) \times {10^4}^g$\\
  NGC 104  & 47Tuc 			&$  4.5^a$ & $  7.4^a$ & $-0.72^a$ & $ 0.09^a$ & $ 2.07^a$ & $ 4.1^a$ & $6.04^g$ & $0.02$ & $5.66^g$ & $0.02$ & $ 9.8^g$ & $ 0.0$ & $		 <  1.5 \times 10^3$\\
  NGC 1851 &       			&$ 12.1^a$ & $ 16.6^a$ & $-1.18^a$ & $ 0.05^a$ & $ 1.86^a$ & $ 1.8^a$ & $5.57^d$ & $0.04$ & $5.25^d$ & $0.04$ & $ 9.3^d$ & $ 0.5$ & $		 <  2.0 \times {10^3}^d$\\
  NGC 1904 &  M79  			&$ 12.9^a$ & $ 18.8^a$ & $-1.60^a$ & $ 0.01^a$ & $ 1.70^a$ & $ 2.4^a$ & $5.15^d$ & $0.03$ & $4.94^d$ & $0.03$ & $ 8.0^d$ & $ 0.5$ & $	( 3.0 \pm  1.0) \times {10^3}^d$\\
  NGC 2808 &       			&$  9.6^a$ & $ 11.1^a$ & $-1.14^a$ & $ 0.12^a$ & $ 1.56^a$ & $ 2.2^a$ & $5.91^c$ & $0.04$ & $5.59^c$ & $0.04$ & $13.4^c$ & $ 0.2$ & $		 <  1.0 \times {10^4}^c$\\
  NGC 5139 &  $\omega$ Cen		&$  5.2^a$ & $  6.4^a$ & $-1.53^a$ & $ 0.17^a$ & $ 1.31^a$ & $ 7.6^a$ & $6.40^k$ & $0.05$ & $5.97^a$ & $0.05$ & $22.0^n$ & $ 4.0$ & $	( 4.7 \pm  1.0) \times {10^4}^k$\\
  NGC 5286 & 				&$ 11.7^a$ & $  8.9^a$ & $-1.69^a$ & $ 0.12^a$ & $ 1.41^a$ & $ 2.5^a$ & $5.45^e$ & $0.02$ & $5.42^e$ & $0.01$ & $ 9.3^e$ & $ 0.4$ & $	( 1.5 \pm  1.1) \times {10^3}^e$\\
  NGC 5694 & 				&$ 35.0^a$ & $ 29.4^a$ & $-1.98^a$ & $ 0.04^a$ & $ 1.89^a$ & $ 4.1^a$ & $5.41^d$ & $0.05$ & $5.09^d$ & $0.05$ & $ 8.8^d$ & $ 0.6$ & $		 <  8.0 \times {10^3}^d$\\
  NGC 5824 & 				&$ 32.1^a$ & $ 25.9^a$ & $-1.91^a$ & $ 0.03^a$ & $ 1.98^a$ & $ 4.2^a$ & $5.65^d$ & $0.03$ & $5.40^d$ & $0.03$ & $11.2^d$ & $ 0.4$ & $		 <  6.0 \times {10^3}^d$\\
  NGC 6093 & M80			&$ 10.0^a$ & $  3.8^a$ & $-1.75^a$ & $ 0.00^a$ & $ 1.68^a$ & $ 1.8^a$ & $5.53^d$ & $0.03$ & $5.17^d$ & $0.03$ & $ 9.3^d$ & $ 0.3$ & $		 <  8.0 \times {10^2}^d$\\
  NGC 6266 & M62			&$  6.8^a$ & $  1.7^a$ & $-1.18^a$ & $ 0.01^a$ & $ 1.71^a$ & $ 1.8^a$ & $5.97^d$ & $0.01$ & $5.57^d$ & $0.01$ & $15.5^d$ & $ 0.5$ & $	( 2.0 \pm  1.0) \times {10^3}^d$\\
  NGC 6388 & 				&$  9.9^a$ & $  3.1^a$ & $-0.55^a$ & $ 0.01^a$ & $ 1.75^a$ & $ 1.5^a$ & $6.04^b$ & $0.08$ & $5.84^b$ & $0.08$ & $18.9^b$ & $ 0.3$ & $	( 1.7 \pm  0.9) \times {10^4}^b$\\
  NGC 6715 & M54			&$ 26.5^a$ & $ 18.9^a$ & $-1.49^a$ & $ 0.06^a$ & $ 2.04^a$ & $ 6.3^a$ & $6.28^g$ & $0.05$ & $6.20^g$ & $0.05$ & $14.2^i$ & $ 1.0$ & $	( 9.4 \pm  5.0) \times {10^3}^l$\\
  NGC 7078 & M15			&$ 10.4^a$ & $ 10.4^a$ & $-2.37^a$ & $ 0.05^a$ & $ 2.29^a$ & $ 3.0^a$ & $5.79^f$ & $0.02$ & $5.59^a$ & $0.02$ & $12.0^f$ & $ 0.0$ & $		 <  4.4 \times {10^3}^f$\\
\noalign{\smallskip}
\hline 
\end{tabular} 
\end{table*}

Given this puzzling behavior of the $\mbh-\sigma$ relation at high masses, the question arises if different laws apply for the low-mass end as well. Attempts to measure lower black-hole masses in dwarf galaxies \citep[e.g.][]{filippenko_2003,barth_2004,xiao_2011} sparsely populate the relation down to a black-hole mass of $\sim 10^5$ but not lower. The lower the black-hole mass, the fainter they appear in radio and X-ray emission and the weaker is the kinematic signature and the smaller the radius of influence. Therefore, detecting black-hole masses lower  than $10^5 \, \msun$ is challenging. Reaching for the mass range which nowadays is assigned to intermediate-mass black holes ($10^2 - 10^5 \, \msun$, IMBHs) requires extending the search towards different stellar systems.

\subsection{Detecting IMBHs in globular clusters}

The possible existence of IMBHs in low-mass stellar systems such as globular clusters (GCs) was first suggested by \cite{silk_1975} while studying X-ray sources in a large sample of globular clusters. They concluded that the observed X-ray fluxes could only be explained by mass accretion onto a $100 - 1000 \, \msun$ central black hole. This discovery triggered a burst of black-hole hunting in globular clusters using X-ray and radio emission but also photometric and kinematic signatures.

Due to the small amount of gas and dust in globular clusters, the accretion efficiency of a potential black hole at the center is expected to be low. This makes the detection of IMBHs at the centers of globular clusters through X-ray and radio emissions very challenging \citep{miller_2002, maccarone_2008}. However, several attempts were made to detect radio and X-ray emission of gas in the central regions and to provide a black-hole mass estimate \citep[e.g.][]{maccarone_2005,ulvestad_2007,bash_2008,cseh_2010}. Recently, \cite{strader_2012} tested several Galactic globular clusters for the presence of possible IMBHs by investigating radio and X-ray emission. They only found upper limits of the order of $10^2 \, \msun$. However, the authors had to make various assumptions about the gas accretion process such as gas distribution, accretion efficiency, and transformation of X-ray fluxes to black-hole masses in order to derive those limits. Besides X-ray and radio emission, the central kinematics in globular clusters can reveal possible IMBHs. 
%However, at the time this method first was proposed forty years ago \citep{wyller_1970,bahcall_1976}, it was limited by the quality of observational datasets. Resolving the gravitational sphere of influence for plausible IMBH masses ($1'' - 2 ''$ for large Galactic globular clusters) requires density and velocity-dispersion measurements at high spatial resolution. Today, with the Hubble Space Telescope (HST) and high spatial resolution ground based integral-field spectrographs, the search for IMBHs is possible.

%\begin{figure}
%  \centering \includegraphics[width=0.5\textwidth]{msig_gal}
%  \caption{The $M_{\bullet} - \sigma$ relation for SMBH in massive galaxies. Overplotted are the best fits to the data from \cite[][purple line]{gultekin_2009} and %\cite[][blue line]{McConnell_2011} with new data of very high-mass black holes.}
%  \label{fig:msig_gal}
%\end{figure}

\subsection{The first IMBH candidates in globular clusters}\label{sec:intro2}

Driven by the results of \cite{silk_1975}, \cite{bahcall_1976} claimed the detection of an IMBH in M15 by measuring its light profile and comparing it to dynamical models. The first claim from kinematic measurements was made by \cite{peterson_1989} who found a strong rise in the velocity dispersion profile of M15. Later, \cite{gebhardt_1997,gebhardt_2000a} and \cite{gerssen_2002} estimated a black-hole mass in M15 of $(3.2 \pm 2.2) \times 10^3 M_{\odot}$ from photometric and kinematic observations. However, after more investigations this cluster no longer appears as a strong IMBH hosting candidate \citep[e.g.][]{dull_1997,baumgardt_2003a,baumgardt_2005,bosch_2006} and the previously detected X-ray sources turned out to be a large number of low-mass X-ray binaries, but new detections of IMBH candidates in other clusters followed. 

Using  integrated light near the center of the M31 cluster G1, \cite{gebhardt_2002,gebhardt_2005} measured the velocity dispersion of this cluster and argued for the presence of a $(1.8 \pm 0.5) \times 10^4 M_{\odot}$ IMBH. Furthermore, X-ray and radio emission were detected at the cluster center, consistent with a black hole of the same mass \citep{pooly_2006, kong_2007,ulvestad_2007}. This result, however, was recently challenged by \cite{miller-jones_2012} who found no radio signature at the center of G1 when repeating the observations.

Another good candidate for hosting an IMBH at its center is the massive globular cluster $\omega$ Centauri (NGC 5139). \cite{noyola_2008, noyola_2010} measured the velocity-dispersion profile with an integral-field unit and used dynamical models to analyze the data. Due to the distinct rise in the velocity-dispersion profile they claim a black-hole mass of $M_{\bullet} = 40~000 \, M_{\odot}$.  The same object was studied by \cite{Avdm_2010} using proper motions from HST images. They found less compelling evidence for a central black hole, but more importantly, they found a location for the center that differs from previous measurements. Both G1 and $\omega$ Centauri have been suggested to be stripped nuclei of dwarf galaxies \citep{freeman_1993, meylan_2001, jalali_2012} and therefore may not be the best representatives of globular clusters. 

%The globular cluster NGC 6388 is also a good candidate for hosting an IMBH. \cite{nora11} detected a rise in its central velocity-dispersion profile which is consistent with a black hole off $\sim$ 20~000 $M_{\odot}$ at its center. Comparison of radio and X-ray emission, however, did not suggest an accreting IMBH. \cite{cseh_2010} found an upper limit of $1500 - 700 \, M_{\odot}$ for the IMBH.

Further evidence for the existence of IMBHs is provided by the discovery of ultra luminous X-ray (ULX) sources at non-nuclear locations in starburst galaxies \citep[e.g.][]{fabbiano_1989, colbert_1999, matsumoto_2001, fabiano_2001}. The brightest of these compact objects (with $L \sim 10^{41} \ \mathrm{erg}\, \mathrm{s}^{-1}$) imply masses larger than $10^3 M_{\odot}$ assuming no beaming of the X-ray emission and accretion at the Eddington limit. Probably one of the best IMBH candidates, the ULX source HLX-1, was discovered in an off-center position of the spiral galaxy ESO243-49 by \citet{farrell_2009} and \citet{godet_2009} and is most likely associated with a group of young stars remaining from the core of a stripped dwarf galaxy \citep[e.g.][]{soria_2010,soria_2013}.

%In order to explain the growth of IMBHs in globular clusters there are two main formation scenarios: a) IMBHs are Population III stellar remnants \citep{madau_2001}, or b) they form in a runaway merging of young stars in sufficiently dense clusters \citep{zwart_2004, gurkan_2004, freitag_2006}. In addition, \cite{miller_2002} presented scenarios for the capture of clusters by their host galaxies and accretion in the galactic disk in order to explain the observed bright X-ray sources.  The understanding of the formation and evolution of IMBHs is crucial for the understanding of the evolution and formation of supermassive black holes. In order to explain the fast formation process of these massive black holes which are observed at very high redshift, i.e. at early times in the Universe, seed black holes are needed \citep{fan_2006}. If IMBHs exist and were formed in globular clusters, they could provide these seeds after being accreted by their host galaxy  \cite[eg.][]{ebisuzaki_2001,tanaka_2009}.

\begin{figure*}
  \centering \includegraphics[width= \textwidth]{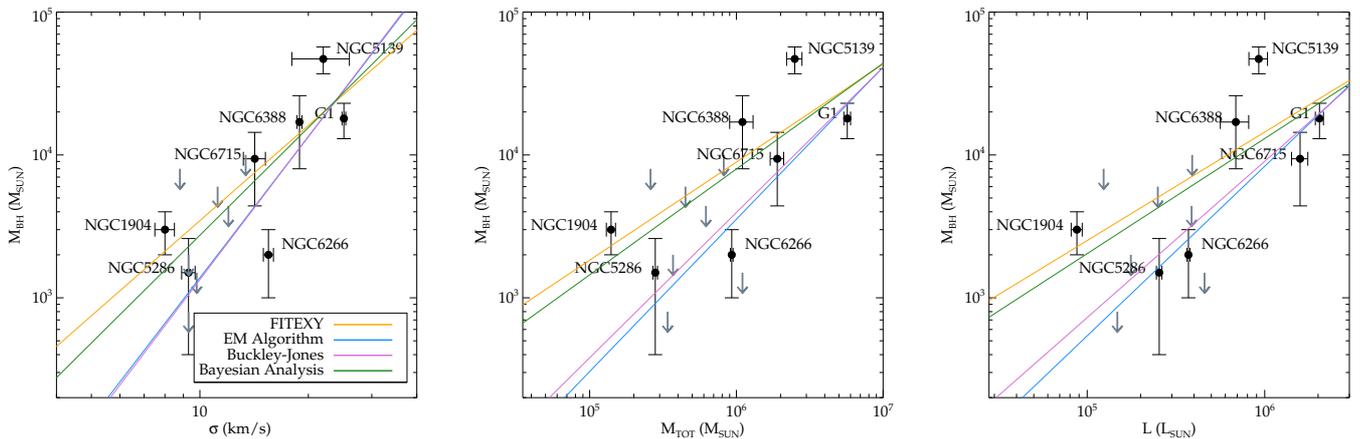}
  \caption{$M_{\bullet} - \sigma$ (left), $M_{\bullet} - M_{TOT}$ (middle) and $M_{\bullet} - L_{TOT}$ (right) relation for globular clusters with IMBHs. The gray arrows indicate upper limits. We compare the slopes of the best-fits from four different methods.}
  \label{fig:major}
\end{figure*}

In this paper we take the latest results for IMBH measurements in globular clusters from the literature and search for correlations with properties of their host cluster. For the first time, enough datapoints on IMBHs are available in order to perform this analysis. Established relations for supermassive black holes and their host galaxies, such as the $M_{\bullet} - \sigma$ relation or the $M_{\bullet} - M_{tot}$, are the focus of this work. Probing the low-mass end of these correlations will provide important information about the origin and growth of supermassive black holes. In Section \ref{sec:sample} we introduce the sample and the data from the literature used in this work. Section \ref{sec:data} describes the methods we apply for determining correlation coefficients and regression parameters and in Section \ref{sec:corr} we discuss the major correlations and their statistical significance. Finally, Section \ref{sec:con} summarizes our work and lists our conclusions.

\section{The sample} \label{sec:sample}

The sample we chose for our analysis consists of 14 Galactic globular clusters for which we have kinematic measurements of their central regions.  The majority of these measurements come from integral-field spectroscopy and was carried out by our group \citep[][Feldmeier et al., 2013 submitted]{nora11,nora12a,nora13}. 

These clusters were observed with the GIRAFFE spectrograph of the FLAMES  \citep[Fiber Large Array Multi Element Spectrograph,][]{pasquini_2002} instrument at the Very Large Telescope (VLT) using the ARGUS mode (Large Integral Field Unit). The velocity-dispersion profile was obtained by combining the spectra in radial bins centered around the adopted photometric center and measuring the broadening of the lines using a non parametric line-of-sight-velocity-distribution fitting algorithm.  For larger radii, the kinematic profiles were completed with radial velocity data from the literature, when available. In addition to the spectroscopic data, HST photometry was used to obtain the star catalogs, the photometric center of the cluster and its surface brightness profile. For each cluster photometry and spectroscopy were combined in order to apply analytic Jeans models to the data. The surface brightness profile was used to obtain a model velocity-dispersion profile which was fit to the data by assuming different black-hole masses and $M/L_V$ profiles. The final black-hole masses were obtained from a $\chi^2$ fit to the kinematic data.  Table \ref{tab:prop} lists all the clusters of the sample and their major properties. 

In four clusters of our sample we found signatures of an intermediate-mass black hole. These clusters are NGC 6388, NGC 6266, NGC 1904 and NGC 5286. All of them show a rise in the velocity dispersion and require black-hole masses between $1.5 \times 10^3 - 2 \times 10^4 \, \msun$. A high radial anisotropy in the center of these clusters would also explain the rise in the velocity-dispersion profile. However, N-body simulations have shown that any anisotropy in the center of a globular cluster would be smoothed out after a couple of relaxation times \citep{nora11}. Since all these clusters are older than at least 5-10 relaxation times, we conclude that the best explanation for the rise of the kinematic profile is indeed a black hole. Nevertheless, the result of the cluster NCG 1904 has to be treated carefully since the rise might occur from a mismatch of outer and inner kinematics. The rest of the clusters in our sample shows less evidence for a central black hole with shallow or even dropping velocity-dispersion profiles. For those we adopt the $1 \sigma$ errors of the measurement as the upper limit.

$\omega$ Centauri, G1 and NGC 6715 are globular clusters with measured black-hole masses not obtained by our group. For $\omega$ Centauri we use the latest result of  \cite{noyola_2010} for the black-hole mass of the IMBH in $\omega$ Centauri of $\sim 47~000 \, M_{\odot}$, which was derived by studying VLT/FLAMES integral-field spectroscopy data of the inner kinematics, similar to the method described above. This value was also confirmed through a comparison with N-body simulations by \citet{jalali_2012}. The globular cluster G1 in M31 was observed with the Space Telescope Imaging Spectrograph (STIS) by \citet{gebhardt_2002} and the velocity-dispersion profile compared with dynamical Schwarzschild models. For our analysis we use their derived mass of $\mbh = ( 1.8 \pm  0.5) \times {10^4} \, \msun$. \citet{ibata_2009} measured the inner kinematics of the globular cluster NGC 6715 (M54) using a combination of VLT/FLAMES MEDUSA and ARGUS data. They found a cusp in the inner velocity-dispersion profile consistent with a $\sim 9~400 \, \msun$ intermediate-mass black hole. Unfortunately they did not provide any uncertainties on this measurement since they could not exclude a high amount of radial anisotropy in the center which would also explain the rise of the profile. With our results obtained in \citet{nora11} however, we are confident that high radial anisotropy is not very likely to reside long in the centers of globular clusters. We therefore adopt the IMBH measurement of \cite{ibata_2009} and adopt a conservative uncertainty of 50\% on the black-hole mass. We stress that high anisotropy in a globular cluster is unlikely for undisturbed clusters, however considering the violent environment of some clusters in the sample (i.e. NGC 6388 and NGC 6266), disk and bulge shocking cannot be excluded. The uncertainty in the black-hole mass measurement might therefore be underestimated in some cases.

The last two clusters included in our sample are M15 and 47 Tuc. Both of them seem to have no IMBHs at their centers. M15, which  was observed with several instruments over the last years, was one of the most popular objects for the study of IMBHs. As a post-core collapse cluster is a bad candidate for hosting an IMBH at its center since a massive black hole would enlarge the core of the cluster and prevent core collapse \citep{baumgardt_2003a, noyola_2011}. Also 47~Tuc is not a good candidate for hosting a massive black hole at its center. \cite{47tuc} obtained HST proper motions for 47~Tuc and found no evidence for a rise in its velocity dispersion profile, i.e. an IMBH.

\begin{table*}
%\tiny
\caption{Correlation parameters and significance calculated through various methods.}           % title of Table
\label{tab:corr}      % is used to refer this table in the text
\centering
\begin{tabular}{lcc|cccc}
\hline \hline
\noalign{\smallskip}
 & \multicolumn{2}{c}{STANDARD STATISTICS} & \multicolumn{4}{|c}{SURVIVAL ANALYSIS} \\
 \noalign{\smallskip}
  & \multicolumn{2}{c}{Kendall} & \multicolumn{2}{|c}{Generalized Kendall} & \multicolumn{2}{c}{Cox Proportional Hazard Model} \\
Relation & $\tau$ & $P$ & $z$ & $P$ & $\chi^2$ & $P$  \\
 \noalign{\smallskip}
\hline
\noalign{\smallskip}
   $M_{\bullet} - \sigma$ & $ 0.79$ & $ 0.96$ & $12.57$ & $ 1.00$ & $ 2.62$ & $ 0.99$\\
  $M_{\bullet} - M_{TOT}$ & $ 0.82$ & $ 0.98$ & $ 5.50$ & $ 0.98$ & $ 2.48$ & $ 0.99$\\
  $M_{\bullet} - L_{TOT}$ & $ 0.71$ & $ 0.93$ & $ 6.45$ & $ 0.99$ & $ 2.40$ & $ 0.98$\\
  $M_{\bullet} - D_{SUN}$ & $-0.04$ & $ 0.06$ & $ 3.03$ & $ 0.92$ & $ 0.40$ & $ 0.31$\\
   $M_{\bullet} - D_{GC}$ & $ 0.21$ & $ 0.36$ & $ 0.01$ & $ 0.09$ & $ 0.27$ & $ 0.21$\\
   $M_{\bullet} - [Fe/H]$ & $ 0.32$ & $ 0.52$ & $ 0.37$ & $ 0.46$ & $ 1.00$ & $ 0.68$\\
        $M_{\bullet} - e$ & $ 0.48$ & $ 0.73$ & $ 1.95$ & $ 0.84$ & $ 1.28$ & $ 0.80$\\
        $M_{\bullet} - c$ & $ 0.11$ & $ 0.18$ & $ 0.96$ & $ 0.67$ & $ 0.40$ & $ 0.31$\\
      $M_{\bullet} - r_h$ & $ 0.54$ & $ 0.78$ & $ 2.34$ & $ 0.87$ & $ 1.60$ & $ 0.89$\\
\hline 
\end{tabular} 
\end{table*}

The final sample of globular clusters and their properties are listed in Table \ref{tab:prop}. The cluster parameters were taken from different sources (see references in Table \ref{tab:prop}) and their black-hole masses from the observations listed above.

\section{Dealing with censored data}\label{sec:data}

With our current dataset we encounter several challenges: 1) The dataset is small. With $n = 14$ we enter statistical regimes ($n<30$) where many statistical approaches, i.e. the Spearman correlation test, are not reliable anymore. 2) The dataset contains not only measurements, but also a large number of upper limits. The statistical method which deals with these so called "censored" datasets is called survival analysis. \cite{Feigelson_1985} and \cite{Isobe_1986} proposed several techniques for applying survival analysis to astronomical datasets. Especially treating bivariate data (a dataset with two variables) with upper limits and determining correlation coefficients and linear regressions is described in detail in these papers. However, survival analysis does not treat uncertainties of the detected measurements. This would bias our result since our dataset 3) also contains large asymmetric uncertainties in both variables. 

We are not aware of a heuristic method which treats all of these caveats properly. Therefore, we will present several attempts to analyze the data by accounting for each problem separately and discussing their differences. The goal is to test for possible correlations between two parameters of the dataset. For those where the correlation is significant we obtain the linear regression parameters $\alpha$, $\beta$ and $\epsilon_0$, where

\begin{equation} y = \alpha + \beta \, x
 \end{equation} \label{eq:lowerlaw}and $\epsilon_0$ indicates the intrinsic scatter i.e., dispersion in $y$ due to the objects themselves rather than to measurement errors. 

\subsection{Partly treating uncertainties}

Since there are no correlation coefficients known which would include uncertainties of the measurements, we are limited to the standard statistical methods. Furthermore, due to our small dataset, we cannot perform a meaningful Spearman correlation test. This leaves us with the the Kendall-$\tau$ rank correlation coefficient as a measurement for the significance of the correlation. The $\tau$ parameter is calculated through the numbers of concordant (ranks of both elements agree) and discordant (ranks of both elements differ) datapairs and the total number of elements in the dataset. If the data $x$ and $y$ are independent from each other, then the correlation coefficient $\tau$ becomes zero. Positive and negative signs of the Kendall-$\tau$ imply a correlation and anticorrelation of the data, respectively. Table \ref{tab:corr} lists the Kendall-$\tau$ and the significance of its deviation from zero (P) for several data combinations as calculated from the IDL routine R\_CORRELATE.

As suggested from Table \ref{tab:corr} and Figure \ref{fig:major}, there are three correlations observed in our dataset: $M_{\bullet} - \sigma$, $M_{\bullet} - M_{tot}$ and $M_{\bullet} - L$ as observed for SMBHs in massive galaxies. In order to compare the correlation properties with measurements from galaxies we perform a power-law fit to the data defined as:

\begin{eqnarray} \log_{10} (M_{\bullet}/M_{\odot}) = & \alpha_{\sigma}+ \beta_{\sigma} \log_{10} [\sigma / (200 \, \rm{km s}^{-1})] \label{eq:sig}\\
\log_{10} (M_{\bullet}/M_{\odot}) =& \alpha_{M} + \beta_{M} \log_{10} [M_{tot} / (10^{11} M_{\odot})]  \label{eq:M}\\
\log_{10} (M_{\bullet}/M_{\odot}) =& \alpha_{L} + \beta_{L} \log_{10} [L_{tot} / (10^{11} L_{\odot})] \label{eq:L}
 \end{eqnarray} \label{eq:lowerlaw}

All three relations are fitted using the algorithm described by \cite{tremaine_2002}, where the quantity 

\begin{eqnarray}
\chi^2 = \sum\limits_{i=1}^{N} \frac{[\log_{10} (M_{\bullet,i}) - \alpha - \beta \log_{10} (\sigma/\sigma_0)]^2}{\epsilon_0^2+\epsilon_{M,i}^2 + \beta^2\epsilon_{\sigma,i}^2}
 \end{eqnarray} \label{eq:chi}is minimized and the 68\% confidence intervals of $\alpha$ and $\beta$ are determined by the range for which $\Delta \chi^2 = \chi^2 - \chi_{min}^2 \leqslant 1$. The uncertainties of $M_{\bullet}$ ($\epsilon_{M}$) and $\sigma$ ($\epsilon_{\sigma}$) are considered asymmetrically in the power-law fit and the internal scatter $\epsilon_0$ is set such that the reduced $\chi^2$, i.e. the $\chi^2$ per degree of freedom, is unity after minimization. $\sigma_0$ is chosen to be $200$ \kms (as in Equation \ref{eq:sig}). For Equation \ref{eq:M} and \ref{eq:L}, $\sigma_0$ becomes $M_0 = 10^{11} M_{\odot}$ and $L_0 = 10^{11} M_{\odot}$, respectively. We use the IDL routine PMFITEXY\footnote{Available at: http://purl.org/mike/mpfitexy} \citep{williams_2010} and extend it in order to account for asymmetric errorbars. The final values of the fit are given in Table \ref{tab:fit} which compares the fit parameters from different methods.

\subsection{Treating upper limits}

The methods developed by  \cite{Feigelson_1985} and \cite{Isobe_1986} combine survival analysis with astronomical applications such as determining correlation coefficients and linear regression parameters. The basic ingredients are the survival function $S(z_i)$ (the probability that the object is not detected until $z_i$) and the hazard function $\lambda(z_i)$ (the instantaneous rate of detection at $z_i$ given that the object is undetected before $z_i$). These functions are used to construct statistical quantities and considering the non-detections. 

A package of Fortran routines called ASURV\footnote{Available at: http://astrostatistics.psu.edu/statcodes/asurv} is available for this purpose. The package contains routines for the determination of non-parametric correlation coefficients such as the Cox Proportional Hazard model \citep[Equation 21 in ][]{Isobe_1986}, the generalized Kendall rank correlation (Equation 28) and linear regression methods such as the EM (Expectation-Maximization) algorithm with different distributions. We apply the routines to our data and list the results in Table \ref{tab:corr} for the correlations and Table \ref{tab:fit} for the linear regression parameters. For the generalized Kendall rank correlation the routine does not calculate $\tau$ itself but the number of standard deviations from zero ($z=\tau/[\rm{Var}(\tau)]^{\frac 1 2}$, where $\rm{Var}(\tau)$ is the variance of $\tau$ under the null hypothesis). From this value, the significance of a correlation between the two variables (P) can be found from a table of the integrated Gaussian distribution. In the case of the Cox Proportional Hazard model (CPH) a $\chi^2$ value is defined in order to test for the significance of a correlation and the  probability is taken from $\chi^2$ distribution tables.

We stress that survival analysis a) assumes detections to be exact measurements, i.e. does not treat uncertainties and b) treats upper limits as absolute values, above (or below) where the data is definitely not detected. Since neither is the case for our data, the results have to be treated with care.

\subsection{Combining uncertainties and upper limits}

In order to treat upper limits and uncertainties in a dataset simultaneously, \cite{kelly_2007} developed a Bayesian method that accounts for measurement errors in linear regression. This method is based on a maximum likelihood approach which is generalized in order to deal with multiple independent variables, nondetections (upper limits) and selection effects. 

The basic approach of the routine (available as IDL routine LINMIX\_ERR in the Astrolib package) is to generate a random sample of the regression parameters drawn from the probability distribution of the parameters given the measured data using a Markov chain Monte Carlo algorithm. For each iteration the regression parameters and the parameters of the prior density are updated until the Markov chain converges to the posterior distribution. The saved parameter values can then be treated as a random draw from the posterior. The final values are derived through the mean or median of the random draw.

For performing the Markov chain, two methods are available: The first one is the Gibbs sampler, which simulates new values of the model parameters and missing data at each iteration conditional on the values of the observed data, current model parameters, and missing data \cite{kelly_2007}. The second method is the Metropolis-Hastings algorithm \citep[e.g.][]{metropolis_1953,hastings_1970} which is used when the selection function is not independent of the dependent variables $y$, the measurement errors are large or the sample size is small. Since this is the case for our sample, we use the Metropolis-Hastings algorithm to determine the regression parameters \cite[for more details we refer to][]{kelly_2007}. In Table \ref{tab:fit} we list the parameters obtained with this method. For further analysis we adopt the results of the Bayesian method as the derived fit.

\section{Discussion}  \label{sec:corr}

In this section we discuss the results listed in Table \ref{tab:fit} and compare the resulting correlations of IMBHs with the existing scaling relations of SMBHs in galaxies.

\begin{figure*}
  \centering \includegraphics[width= 0.33 \textwidth]{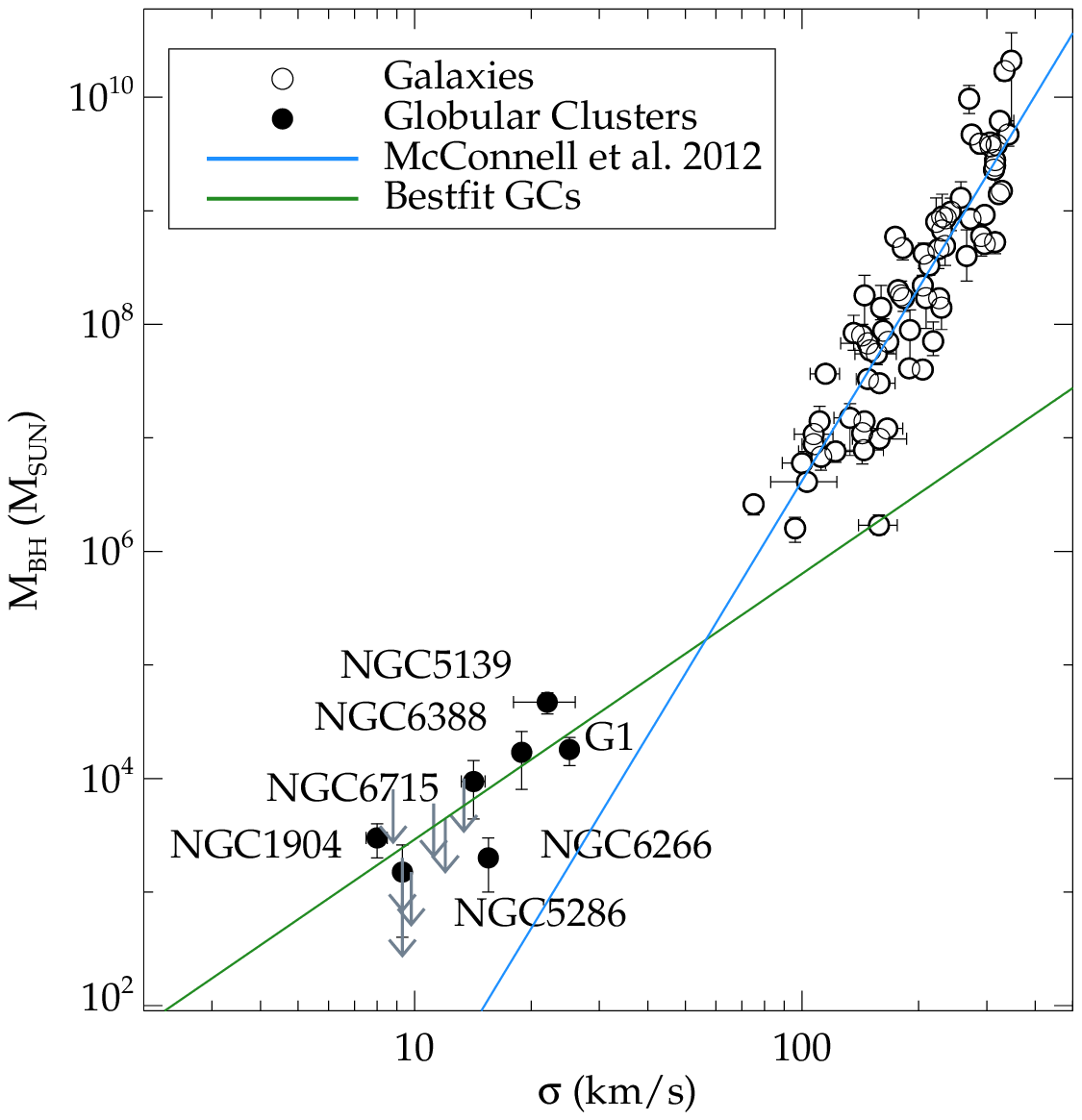}
  \centering \includegraphics[width= 0.33 \textwidth]{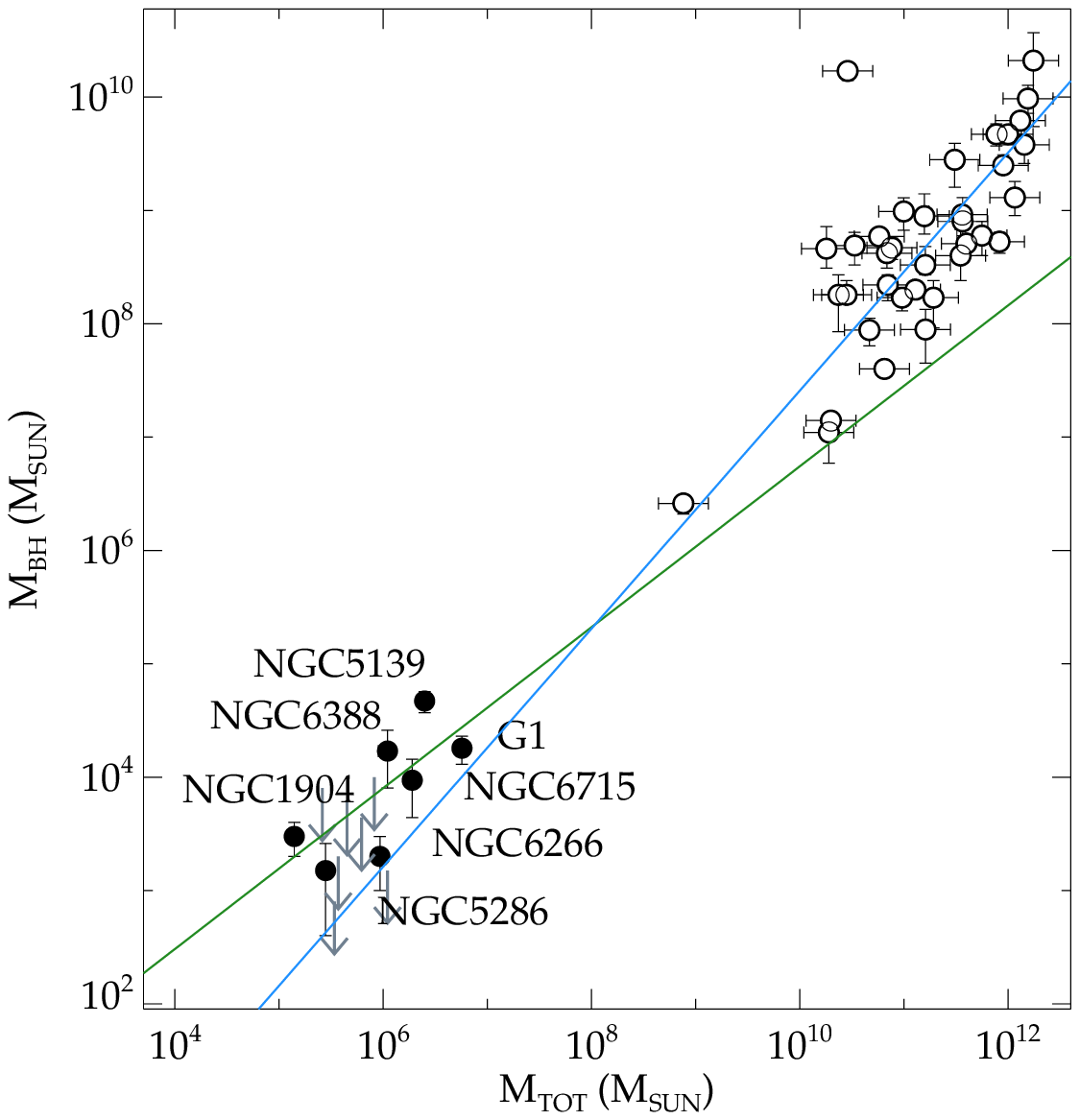}
  \centering \includegraphics[width= 0.33 \textwidth]{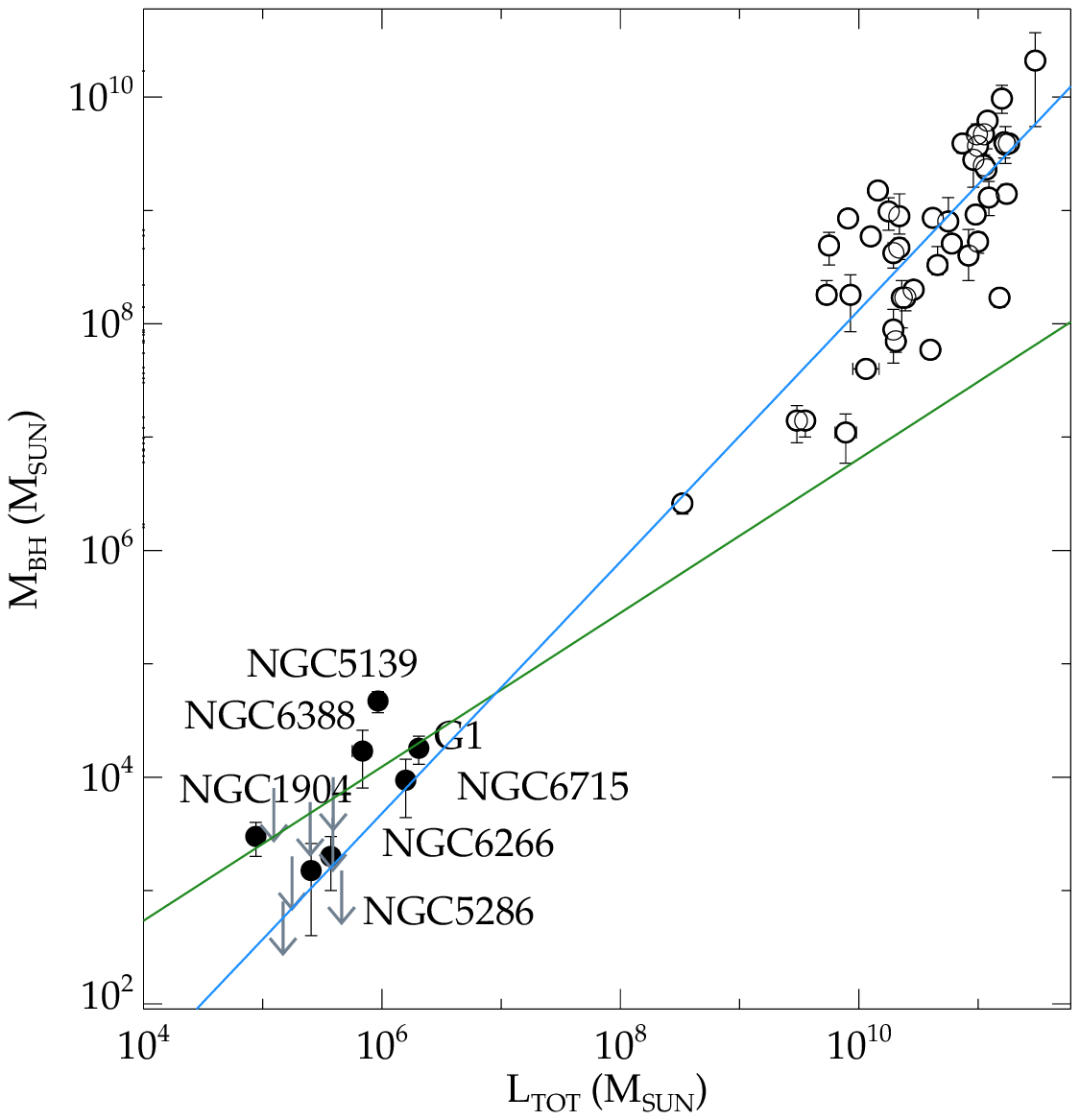}
  \caption{$M_{\bullet} - \sigma$,  $M_{\bullet} - M_{tot}$, and $M_{\bullet} - L_{tot}$  relations of IMBHs and SMBHs in comparison. The slope of the best fit to the GCs (green line) is a factor of two smaller than the slope for the SMBHs in galaxies (blue line) for $M_{\bullet} - \sigma$, but very similar for the remaining two correlations. }
  \label{fig:msig}
\end{figure*}

\subsection{The three major scaling relations}

The first correlations we study are the ones observed in galaxies. Comparing these scaling relations with the ones obtained for globular clusters might shed light onto their origin and the evolution of globular clusters. Therefore, we first examine the relations $M_{\bullet} - \sigma$, $M_{\bullet} - M_{tot}$, and $M_{\bullet} - L$. We test for the significance of a correlation by applying the methods described in the previous section. Table \ref{tab:corr} lists the results of the Kendall $\tau$ test from standard statistics compared to the generalized Kendall $\tau$ and the results from the CPH model from survival analysis. The value P gives the probability of an existing correlation between the two values and shows high significance ($> 90\%$) for all three correlations with all methods. The highest probability is found for the $M_{\bullet} - \sigma$ relation when using the generalized Kendall $\tau$, but when regarding only the standard statistics, the $M_{\bullet} - M_{tot}$ is given the largest significance. It is also the correlation with the most stable results over all three methods. 

The existence of these correlations agrees with previous measurements of SMBHs in galaxies, where the same major scaling relations where found. We apply a linear regression to all three relations, using the methods described in the previous sections. In Figure \ref{fig:msig} we show the three scaling relations, overplotted by the best-fits from the Bayesian method (green line in Figure \ref{fig:major}). For the $M_{\bullet} - \sigma$ relation, all parameters of the linear regression from the different methods agree within their errorbars.  The parameters derived from the different survival analysis methods are almost identical. In general, the plots show that the values of the survival analysis do not differ much from each other for all correlations. The same is true for the FITEXY method and the Bayesian approach, which shows that the errorbars have a larger effect on the fit than the upper limits. 

Comparing these values with slopes obtained for supermassive black holes in galaxies (Figure \ref{fig:msig}, blue line) yields a significant  difference for the slope of the correlation between galaxies and globular clusters in the $M_{\bullet} - \sigma$ relation. The most recent slope obtained from \cite{mcconnell_2012} is $\beta_{\sigma} = 5.64 \pm 0.32$, a factor of two higher than the value that we measured for the globular clusters. This is also visible in Figure \ref{fig:msig} where the values for the galaxies and the globular clusters are overplotted. The  $M_{\bullet} - \sigma$ relation seems to be curved upwards when reaching the very high-mass black holes and becomes more shallow when reaching the low-mass range.

Also for the $M_{\bullet} - M_{tot}$ and $M_{\bullet} - L_{tot}$ correlation all IMBH masses lie above the correlation found for galaxies but, as shown in Figure \ref{fig:msig}, the slopes of the correlations do not differ as much as they do for the $M_{\bullet} - \sigma$ relation. In fact, with $\beta = 1.11  \pm 0.13$ for $M_{\bullet} - L_{tot}$ and $\beta = 1.05  \pm 0.11$ for $M_{\bullet} - L_{tot}$, the values obtained by \cite{mcconnell_2012} agree with the values from Table \ref{tab:fit} within their uncertainties. %\textbf{The IMBH mass measurements coincide with upper limits of IMBH masses in nuclear star clusters on the $M_{\bullet} - \sigma$ relation \citep{neumayer_2012}.}

One possible explanation for the difference in the scaling relations could be a strong mass-loss of globular clusters in the early stages of their evolution. Especially for $\omega$ Centauri, NGC 1851 and G1 this could have had a large effect as these objects were suggested to be cores of stripped dwarf galaxies \citep{freeman_1993, meylan_2001, jalali_2012,sollima_2012}. A higher mass in the early stage of evolution of these objects would have shifted the points in all three plots to the right and therefore closer to the correlations observed for galaxies. The strongest difference in slopes is shown in the $M_{\bullet} - \sigma$ relation. Expansion due to mass loss and relaxation would lead to an decrease of the velocity dispersion and could explain the different effect on $M_{tot}$ and $\sigma$. 

Another explanation would be a different mass-radius relation for galaxies and globular clusters which complicates the comparison of the scaling relations as the velocity dispersion is measured from an effective radius in both systems. Since globular clusters and galaxies are different objects, formed in different environments and processes, this explanation seems reasonable and is discussed in \citet{graham_2011} and references therein. The fact that our relations match with the results found for nuclear star clusters \citep{graham_2012,neumayer_2012}, which are found to exhibit shallow slopes, similar to globular clusters, supports this theory.  Furthermore, clusters in our sample with upper limits still have a probability of hosting smaller IMBHs not detected up to date. This would bias all the relations towards high IMBH masses and could partly explain the difference in slopes. In fact, this is indicated by the steeper slopes from the survival analysis (where we consider only upper limits but no errorbars) as shown in Figure \ref{fig:major}.

\begin{figure*}
  \centering \includegraphics[width= \textwidth]{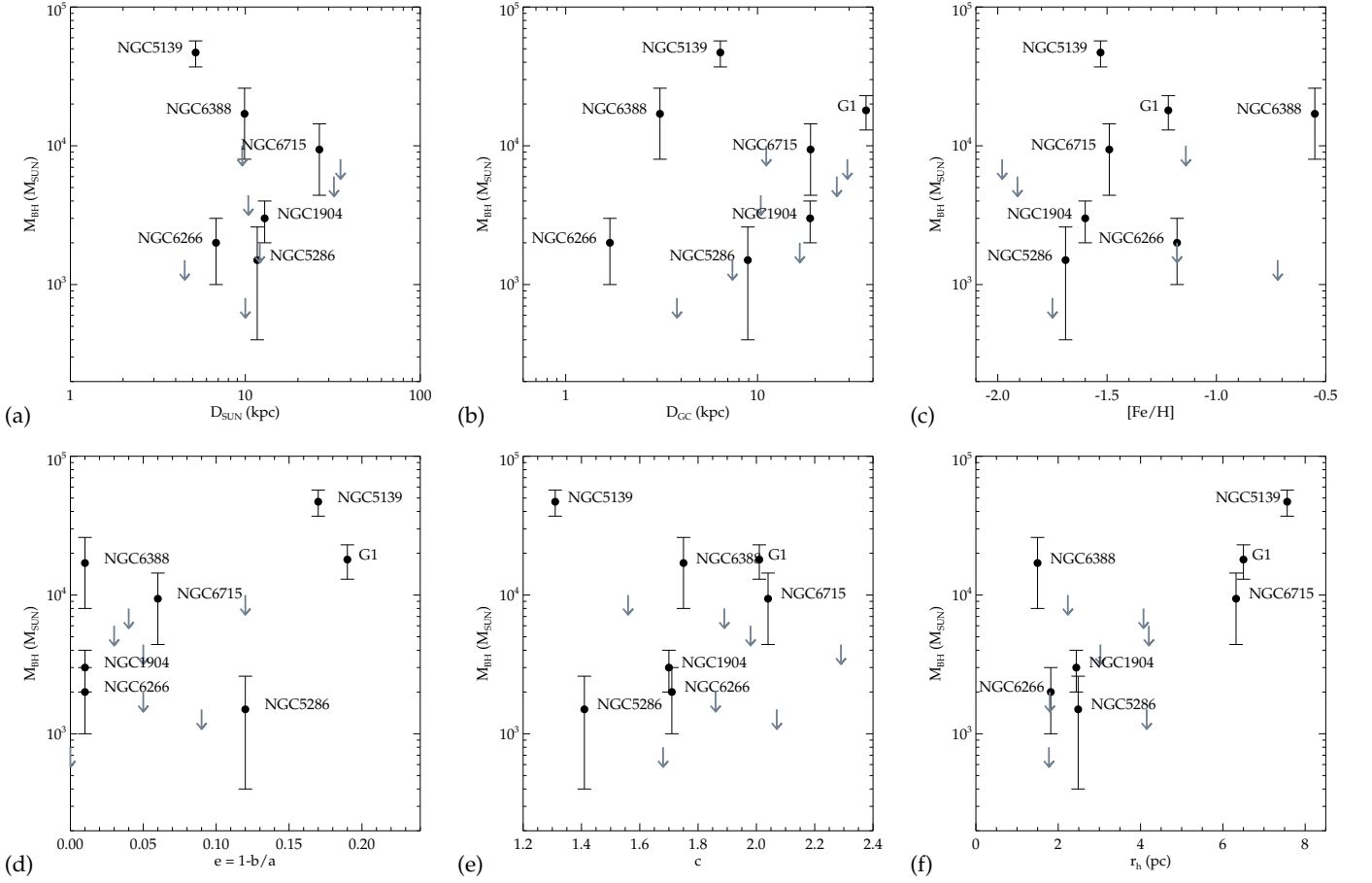}
  \caption{IMBH masses as a function of several cluster properties such as a) the distance of the cluster to the sun ($D_{SUN}$), b) the distance from the Galactic center ($D_{GC}$), c) the metallicity ([Fe/H]), d) the ellipticity ($e = 1- b/a$), e) the concentration parameter (c), and f) the half-mass radius ($r_h$)}
  \label{fig:others}
\end{figure*}

\subsection{Are there further correlations?}

The three major scaling relations discussed above are known for galaxies that host SMBHs at their centers. Even for the cases where the slopes of the IMBH correlations agree with the SMBH, it does not imply a causal connection. Formation scenarios and co-evolution of host system and black hole are assumed to be quite different for galaxies and globular clusters. For this reason it is crucial to search for fundamental correlations of cluster-specific values with the mass of the IMBH. This has never been done before. This section is dedicated to describe the choice of parameters and the results on their correlation significance.

\subsubsection{Distances}

The first parameter we check is the distance to the sun ($D_{SUN}$). No correlation is expected with this parameter, but it can be used to exclude observational biases that might be introduced to the measurements. The further away the object, the larger one would expect the uncertainty on the black-hole mass to be. Panel a) of Figure \ref{fig:others} shows the black-hole mass as a function of distance for all Galactic globular clusters in our sample. We exclude G1 in the plot (but not in the analysis) as it is the only cluster outside the Milky Way and its large distance would distort the figure. From the plot no evident correlation between $D_{SUN}$ and $M_{\bullet}$ can be found, but the correlation coefficients in Table \ref{tab:corr} show a large spread reaching from a slight anti-correlation for the standard Kendall $\tau$ to a significant correlation when using survival analysis. The reason for this mismatch might arise from the outlier position of G1 compared to the other distances. 

As another parameter we study the distance of the cluster to the Galactic center (in case of G1 to the center of M31), i.e. the position of the cluster in its host galaxy ($D_{GC}$). Finding a correlation or anticorrelation with galactocentric distance, could open speculations about a connection between environment and black-hole formation, although the clusters are most probably not at the same distance where they were formed. Panel b) and the results in Table \ref{tab:corr} do not indicate any correlation with $D_{GC}$.

\subsubsection{Metallicity}

In Panel c) we correlate the black-hole mass with the metallicity of the cluster. A cluster with a low metallicity might be more likely to form a massive black hole through runaway merging of massive stars due to the lack of strong stellar winds at low metallicities \citep[e.g.][]{vanbeveren_2009}. However, according to Figure \ref{fig:others}.c  and the values in Table \ref{tab:corr}, this does not seem to be the case.

\subsubsection{Morphology}

The lower Panels in Figure \ref{fig:others} depict the black-hole mass as a function of the three structural parameters: ellipticity ($e$), concentration ($c$) and half-mass radius ($r_h$). The reason why we test for correlations in these parameters lies in the possible connection between morphology and the cluster being the nucleus of an accreted dwarf galaxy. For example, a high ellipticity would speak against globular clusters which are assumed to be spherical and dynamically relaxed systems and for a remnant of a dwarf galaxy where aspherical shapes are more common.  In the same context, an IMBH in a dwarf galaxy is thought to be more massive than in a globular cluster since more mass was available at its formation. 

Furthermore it has been shown that a black hole prevents core collapse and extends the inner regions of the cluster \citep[e.g.][]{baumgardt_2005}. Therefore, a correlation of the central concentration with black-hole mass would be expected. Also clusters with large half-mass radii have been suggested of being good candidates of hosting IMBHs at their centers.  The highest correlation significance is found for the half-mass radius but also for the ellipticity  a $1-2 \, \sigma$ detection of correlation is present. For the concentration, the values of the different methods show large variations ranging from $18 \%$ to $67 \%$. In summary, the structural parameters and the black-hole mass show signs of correlations. However, the detection significance is low and needs to be confirmed with a larger data set.

\begin{table}
%\tiny
\caption{Best-fit parameters for the three major scaling relations obtained form different fitting routines.}           % title of Table
\label{tab:fit}      % is used to refer this table in the text
\centering
\begin{tabular}{lcccc}
\hline \hline
\noalign{\smallskip}
 & $M_{\bullet} - \sigma$& $M_{\bullet} - M_{TOT}$  & $M_{\bullet} - L_{TOT}$ \\
 \noalign{\smallskip}
\hline
\noalign{\smallskip}
 & \multicolumn{3}{c}{---FITEXY---}\\
\noalign{\smallskip}
$\alpha$ 	&$ 6.42\pm 0.77$&$ 7.38\pm 1.40$&$ 7.96\pm 1.81$\\
$\beta$		&$ 2.21\pm 0.69$&$ 0.69\pm 0.28$&$ 0.76\pm 0.34$\\
$\epsilon$	&$ 0.25$&$ 0.33$&$ 0.35$\\
\noalign{\smallskip}
 & \multicolumn{3}{c}{---SURVIVAL ANALYSIS---}\\
  & \multicolumn{3}{c}{EM Algorithm}\\
\noalign{\smallskip}
$\alpha$	&$ 7.41\pm 0.83$&$ 8.63\pm 2.03$&$ 9.84\pm 2.10$\\
$\beta$	&$ 3.28\pm 0.72$&$ 1.01\pm 0.34$&$ 1.18\pm 0.37$\\
$\sigma$	&$ 0.36$&$ 0.49$&$ 0.47$\\
  & \multicolumn{3}{c}{Buckley-Jones Method}\\
$\alpha$	&$ 7.45$&$ 8.42$&$ 9.47$\\
$\beta$	&$ 3.32\pm 0.92$&$ 0.96\pm 0.32$&$ 1.10\pm 0.40$\\
$\sigma$	&$ 0.42$&$ 0.43$&$ 0.46$\\
\noalign{\smallskip}
 & \multicolumn{3}{c}{---BAYESIAN ANALYSIS---}\\
\noalign{\smallskip}
$\alpha$	&$ 6.51\pm 1.94$&$ 7.45\pm 3.53$&$ 7.49\pm 6.08$\\
$\beta$ &$ 2.34\pm 1.63$&$ 0.71\pm 0.59$&$ 0.68\pm 1.05$\\

\hline 
\end{tabular} 
\end{table}

\section{Summary and Conclusions}\label{sec:con}

We collected data from the literature from globular clusters that were examined for the existence of a possible IMBH at their center using stellar dynamics. Our sample consists of 14 globular clusters with kinematically measured black-hole masses: six of which are thought to host IMBHs, the remaining ones having only upper limits on black-hole masses. In order to take uncertainties and upper limits into account, we use different methods to derive the correlation coefficients and linear regression parameters. We plot the masses of the central IMBH versus several properties of the host clusters, such as total mass, total luminosity, and velocity dispersion in order to verify existing correlations which are found in galaxies, but also with the attempt to find new correlations between black-hole mass and cluster properties. We find that the three main correlations which are observed in galaxies ($M_{\bullet} - \sigma$, $M_{\bullet} - M_{tot}$, and $M_{\bullet} - L_{tot}$) also hold for IMBHs. The slope of the $M_{\bullet} - \sigma$ relations differs by a factor of two from the fits made to the galaxy sample while the remaining correlations are more similar to the ones observed for galaxies.

Furthermore, we test for possible correlations of the IMBH mass and other properties of the globular cluster such as distance, metallicity, structural properties and size. We find no evidence for a correlation as strong as the $M_{\bullet} - \sigma$ relation, but we find a trend of black-hole mass with the cluster size, i.e. the half-mass radius $r_h$ and its ellipticity. This is reasonable since the central black hole leads the cluster to expand by accelerating stars in its vicinity. 

We assume in the following that the $\mbh - \sigma$ relation is tight, has a physical origin and extends to the lowest masses. These assumptions have not rigorously been demonstrated to date. We assume further that the IMBHs in globular clusters formed at high redshifts and did not evolve much since then, unlike the globular clusters themselves which will have suffered mass loss during their evolution. With these assumptions, a proposed explanation for this behavior is the process of stripping which might have occurred to these stellar systems when accreted to the Milky Way. If they were more massive and luminous in the past they would have fit to the correlation of the galaxies instead of being shifted to the low-mass end. Especially, these stripping scenarios have already been suggested for clusters like G1 and $\omega$ Centauri, as mentioned in Section \ref{sec:intro2}. In particular, given the fact that clusters most probably lost most of their mass during their lifetime and galaxies gained mass during their merging history, it would be very surprising to find both objects on the same scaling relation.  

The fact that the $\mbh - \sigma$ shows larger discrepancies when comparing to the relation of galaxies than $M_{\bullet} - M_{tot}$ and $M_{\bullet} - L_{tot}$ might arise in either different mass-radius relations of these very different black-hole host systems as well as the expansion driven decrease of the velocity dispersion in globular clusters. This is supported by the fact that the scaling relations agree with the upper limits for nuclear clusters, stellar systems more similar to globular clusters. N-body simulations and semi-analytic expansion models are needed in order to quantitatively study these theories. We note that also a bias towards higher black-hole masses due to detection limits can not be excluded.

For the future it is desired to extend this sample and confirm the reported correlations and their slopes. With integral-field units combined with adaptive optics, the search for IMBHs can be extended to extragalactic sources. Nevertheless, a complete search of Galactic globular clusters will provide a larger sample and  help to further constrain critical observables which hint towards the presence of an IMBH in the center of a globular cluster.

\begin{acknowledgements}
This research was supported by the DFG cluster of excellence Origin and Structure of the Universe (www.universe-cluster.de). H.B. acknowledges support from the Australian Research Council through Future Fellowship grant FT0991052. N.L. thanks Eric Feigelson, Brandon Kelly and Michael Williams for their friendly support and for providing the essential software used in this work. We thank the referee for constructive comments and suggestions that helped to improve this manuscript.
\end{acknowledgements}

\bibliographystyle{aa}
\bibliography{ref}

\end{document}